\documentclass[english]{article}%
\usepackage{amssymb}
\usepackage{fontenc}
\usepackage[latin1]{inputenc}
\usepackage{amsmath}
\usepackage{fontenc}
\usepackage{babel}%
\usepackage{amsfonts}%
\usepackage{graphicx}
\providecommand{\U}[1]{\protect\rule{.1in}{.1in}}
 
\begin{document}

\title{{\large Coherent states of non-relativistic electron in magnetic -- solenoid
field}}
\author{V. G. Bagrov\thanks{Department of Physics, Tomsk State University, 634050,
Tomsk, Russia. Tomsk Institute of High Current Electronics, SB RAS, 634034
Tomsk, Russia; e-mail: bagrov@phys.tsu.ru} , S. P. Gavrilov\thanks{Institute
of Physics, University of São Paulo, Brazil; On leave from Department of
general and experimental physics, Herzen State Pedagogical University of
Russia, Moyka emb. 48, 191186 St. Petersburg, Russia; e-mail:
gavrilovsergeyp@yahoo.com}, D. M. Gitman\thanks{Institute of Physics,
University of São Paulo, CP 66318, CEP 05315-970 São Paulo, SP, Brazil;
e-mail: gitman@dfn.if.usp.br}, D. P. Meira Filho\thanks{Institute of Physics,
University of São Paulo, Brazil; e-mail: dmeira@dfn.if.usp.br}}
\maketitle

\begin{abstract}
In the present work we construct coherent states in the magnetic-solenoid
field, which is a superposition of the Aharonov-Bohm field and a collinear
uniform magnetic field. In the problem under consideration there are two kind
of coherent states, those which correspond classical trajectories which
embrace the solenoid and those which do not. Constructed coherent states
reproduce exactly classical trajectories, they maintain their form under the
time evolution, and form a complete set of functions, which can be useful in
semiclassical calculations. In the absence of the solenoid filed these states
are reduced to well known in the case of a uniform magnetic field
Malkin-Man'ko coherent states.

PACS numbers: 03.65.Ge, 03.65.Sq

\end{abstract}

\section{Introduction}

It is well-known that the study of the Aharonov-Bohm (AB) effect\ began on the
base of exact wave functions of an electron in the field of an infinitely long
and infinitesimally thin solenoid \cite{AB}.\ Such functions allow one to
analyze a nontrivial influence of the AB solenoid (ABS) on scattering of free
electron, which may give a new interpretation of electromagnetic potentials in
quantum theory. Physically it is clear, that in such a scattering, the
electron is subjected to the action of the AB field for a short finite time.
However, there exist a possibility to consider bound states of the electron in
which it is affected by the AB field for the infinite time. Such bound states
exist in the so-called magnetic-solenoid field (MSF), which is a superposition
of the AB field and a collinear uniform magnetic field. The non-relativistic
and relativistic wave functions of an electron in the MSF were studied in
\cite{L83,RelExSol,ExnSV02,GSTV09,GavGSV04}. In particular, they were used to
describe AB effect in cyclotron and synchrotron radiations \cite{Synch}. We
believe that such bound states of an electron in the MSF open new
possibilities the study of the AB effect. One of the important questions is
the construction and study of semiclassical (coherent) states in the MSF (the
importance and advantage of coherent states in quantum theory is well-known
\cite{CSQT72}). Having such states in hands one can try to answer an important
question: to what extend the AB effect is of a pure quantum nature. In a sense
constructing semiclassical states is a complimentary task to the path integral
construction. The latter problem is completely open in the case of the
particle in the magnetic-solenoid field. One ought to say that some attempts
to construct semiclassical states in the MSF were made in the works
\cite{CS-old}. However, one ought to accept that states constructed in these
works have some features that does not allow one to interpret them as
semiclassical and coherent states. For example, some mean values calculated in
such states do not move along classical trajectories. In the present work we
succeeded to construct another kind of semiclassical states in the MSF which
can be really treated as coherent states. The progress is related to a
nontrivial observation that in the problem under consideration there are two
kind of coherent states those which correspond classical trajectories which
embrace the solenoid and those which do not. Constructed coherent states
reproduce classical trajectories in the semiclassical limit, they maintain
their form under the time evolution, and form a complete set of functions,
which can be useful in semiclassical calculations. In the absence of the AB
field these states are reduced to well known in the case of a uniform magnetic
field Malkin-Man'ko coherent states \cite{MalMa68}.

We consider the non-relativistic motion of an electron with charge $q=-e$,
$e>0$, and mass $M$ in the MSF $\mathbf{B}=\left(  B_{x},B_{y},B_{z}\right)
,$%
\begin{equation}
B_{x}=B_{y}=0,\ B_{z}=B+\Phi\delta\left(  x\right)  \delta\left(  y\right)
=B+\frac{\Phi}{\pi r}\delta\left(  r\right)  ,\,\,\label{2.1}%
\end{equation}
which is a collinear superposition of a constant uniform magnetic field $B$
directed along the axis $z$ ($B>0$) and the AB field (field of an infinitely
long and infinitesimally thin solenoid) with a finite constant internal
magnetic flux $\Phi$. We use Cartesian coordinates $x$, $y$, $z$, as well as
cylindrical coordinates $r,\varphi$, such that$\ x=r\cos\varphi$,
$y=r\sin\varphi$, and $r^{2}=x^{2}+y^{2}$. The field (\ref{2.1}) can be
described by the vector potential $\mathbf{A}=\left(  A_{x},A_{y}%
,A_{z}\right)  ,$%
\begin{equation}
A_{x}=-y\left(  \frac{\Phi}{2\pi r^{2}}+\frac{B}{2}\right)  ,\ A_{y}=x\left(
\frac{\Phi}{2\pi r^{2}}+\frac{B}{2}\right)  ,\;A_{z}=0.\label{2.2}%
\end{equation}

Classical motion of the electron in the MSF is governed by the Hamiltonian
$H=\mathbf{P}^{2}/2M$,$\ \mathbf{P}=\mathbf{p}-\frac{q}{c}\mathbf{A}%
$\textbf{,}$\mathbf{\ }$where $\mathbf{p}$ and $\mathbf{P}$ are the
generalized and kinetic momentum, respectively. Trajectories that do not
intersect the solenoid have the form:%
\begin{equation}
x=x_{0}+R\cos\psi,\ \ y=y_{0}+R\sin\psi,\;z=\frac{p_{z}}{M}t+z_{0}%
;\;\psi=\omega t+\psi_{0},\;\omega=\frac{eB}{Mc},\label{l2}%
\end{equation}
where $x_{0},\,y_{0},\,z_{0},\,p_{\,z},\,R,$ and$\,\psi_{0}$ are integration
constants. Eqs. (\ref{l2}) imply%
\begin{align}
& \ (x-x_{0})^{2}+(y-y_{0})^{2}=R^{2},\ \ x_{0}=R_{c}\cos\alpha,\ \ y_{0}%
=R_{c}\sin\alpha,\nonumber\\
& r^{2}=R^{2}+R_{c}^{2}+2RR_{c}\cos(\psi-\alpha),\;R_{c}=\sqrt{x_{0}^{2}%
+y_{0}^{2}}\ .\label{l3}%
\end{align}
The projection of particle trajectories on the $xy$-plane are circles.
Particle images on the $xy$-plane are rotating with the synchrotron frequency
$\omega$. For an observer which is placed near the solenoid with $z>0,$ the
rotation is anticlockwise. The particle has a constant velocity $p_{z}/M$
along the axis $z$ . Since the electron freely propagates on the $z$ axis,
only motion in the perpendicular plane $z=0$ is nontrivial; this will be
examined below. Denoting by $r_{max}$ the maximal possible moving off and by
$r_{min}$ the minimal possible moving off of the particle from the $z$-axis,
we obtain from (\ref{l3}) $r_{max}=R+R_{c}$,$\ r_{min}=|R-R_{c}|$. It follows
from (\ref{l2}) that%
\begin{align}
P_{x}  & =-M\omega\,R\sin\psi=-M\omega\left(  y-y_{0}\right)  ,\nonumber\\
P_{y}  & =M\omega R\cos\psi=\,M\omega\left(  x-x_{0}\right)  ,\ \ \mathbf{P}%
_{\bot}^{2}=P_{x}^{2}+P_{y}^{2}=(M\omega\,R)^{2}.\label{l5}%
\end{align}
The energy $E$ of the particle rotation reads: $E=\mathbf{P}_{\bot}^{2}/2M$
then the radius $R$\ can be expressed via the energy $E$\ as follows%
\begin{equation}
R^{2}=\frac{2E}{M\omega^{2}}.\label{17a}%
\end{equation}
By the help of (\ref{l5}), one can calculate angular momentum projection
$L_{z}$,
\begin{equation}
L_{z}=xp_{y}-yp_{x}=\frac{M\omega}{2}(R^{2}-R_{c}^{2})-\frac{e\Phi}{2\pi
c}\ .\label{17}%
\end{equation}

\begin{figure}[ptb]%
\centering
\includegraphics[
height=2.6394in,
width=2.7665in
]%
{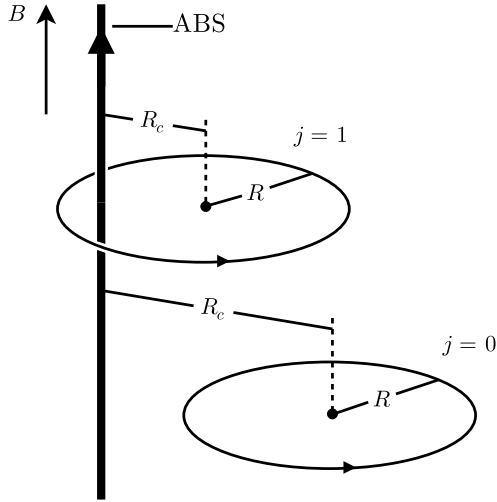}%
\caption{Two types of trajectories in MSF}%
\label{twotypes}%
\end{figure}

The presence of ABS breaks the translational symmetry in the $xy$-plane, which
on the classical level has only a topological effect, there appear two types
of trajectories, we label them by an index $j=0,1$ in what follows. On the
classical level $j=1$ corresponds to $(R^{2}-R_{c}^{2})>0$ and $j=0$
corresponds to $(R^{2}-R_{c}^{2})<0,$ see Fig. \ref{twotypes}.%

Already in classical theory, it is convenient, to introduce dimensionless
complex quantities $a_{1}$ and $a_{2}$ (containing $\hbar$) as follows:
\begin{equation}
a_{1}=\frac{-iP_{x}-P_{y}}{\sqrt{2\hbar\,M\omega}}=-\sqrt{\frac{M\omega
}{2\hbar}}Re^{-i\psi},\ \ a_{2}=\frac{M\omega\,(x+iy)+iP_{x}-P_{y}}%
{\sqrt{2\hbar\,M\omega}}=\sqrt{\frac{M\omega}{2\hbar}}R_{c}e^{i\alpha
}.\label{l9}%
\end{equation}
One can see that $a_{1}\exp(i\psi)$ and $a_{2}$ are complex (dependent)
integrals of motion. One can write that%
\begin{align}
& R^{2}=\frac{2\hbar}{M\omega}a_{1}^{\ast}a_{1},\;R_{c}^{2}=\frac{2\hbar
}{M\omega}a_{2}^{\ast}a_{2},\ x+iy=\sqrt{\frac{2\hbar}{M\omega}}\left(
a_{2}-a_{1}^{\ast}\right)  ,\label{20}\\
& E=\omega\hbar a_{1}^{\ast}a_{1},\ L_{z}=\hbar\left(  a_{1}^{\ast}a_{1}%
-a_{2}^{\ast}a_{2}\right)  -\frac{e\Phi}{2\pi c}\ .\label{21}%
\end{align}

\section{Stationary states}

The quantum behavior of the electron in the field (\ref{2.1}) is determined by
the Schrödinger equation with the Hamiltonian
\begin{align}
\hat{H}  & =\hat{H}_{\bot}+\hat{p}_{z}^{2}/2M,\;\;\hat{H}_{\bot}=\left(
\hat{P}_{x}^{2}+\hat{P}_{y}^{2}\right)  /2M,\;\nonumber\\
\hat{P}_{x}  & =\hat{p}_{x}+\frac{e}{c}A_{x},\ \ \hat{P}_{y}=\hat{p}_{y}%
+\frac{e}{c}A_{y},\ \ \hat{p}_{x}=-i\hbar\partial_{x},\ \ \hat{p}_{y}%
=-i\hbar\partial_{y},\;\hat{p}_{z}=-i\hbar\partial_{z},\label{H}%
\end{align}
where $\hat{H}_{\bot}$ determines the nontrivial behavior on the $xy$-plane.
It is convenient to present magnetic flux $\Phi$ in eq. (\ref{2.2}) as
$\Phi=\left(  l_{0}+\mu\right)  \Phi_{0},$ where $l_{0}$ is integer, and
$0\leq\mu<1$ and $\Phi_{0}=2\pi c\hbar/e$ is Dirac's fundamental unit of
magnetic flux. Mantissa of the magnetic flux $\mu$ determines, in fact, all
the quantum effects due to the presence of the AB field. Stationary states of
the non-relativistic electron in the MSF were first described in \cite{L83}.
The corresponding radial functions were taken regular at $r\rightarrow0$, they
correspond to a most natural self-adjoint extension (with a domain
$D_{H_{\bot}}$) of the differential symmetric operator $\hat{H}_{\bot}$.
Considering a regularized case of a finite-radius solenoid one can demonstrate
that the zero-radius limit yields such an extension, see \cite{GavGSV04}.
Further, we consider only such an extension (all possible self-adjoint
extensions of $\hat{H}_{\bot}$ were constructed in (\cite{ExnSV02,GSTV09}).
Operator $\hat{L}_{z}=x\hat{p}_{y}-y\hat{p}_{x}$ is self-adjoint on
$D_{H_{\bot}}$ and commutes with the self-adjoint Hamiltonian $\hat{H}_{\bot}%
$. One can find two types ($j=0,1)$ of common eigenfunctions of both
operators
\begin{align}
\hat{H}_{\bot}\Psi_{n_{1},\,n_{2}}^{(j)}\left(  t,r,\varphi\right)   &
=\mathcal{E}_{n_{1}}\Psi_{n_{1},\,n_{2}}^{(j)}\left(  t,r,\varphi\right)
,\ \ \mathcal{E}_{n_{1}}=\hbar\omega\left(  n_{1}+1/2\right)  ,\nonumber\\
\hat{L}_{z}\Psi_{n_{1},\,n_{2}}^{(j)}\left(  t,r,\varphi\right)   & =L_{z}%
\Psi_{n_{1},\,n_{2}}^{(j)}\left(  t,r,\varphi\right)  ,\ \ L_{z}=\hbar\left(
l-l_{0}\right)  .\label{2.15}%
\end{align}

The eigenfunctions have the form%
\begin{align}
&  \Psi_{n_{1},\,n_{2}}^{(j)}\left(  t,r,\varphi\right)  =\exp\left(
-\frac{i}{\hbar}\mathcal{E}_{n_{1}}t\right)  \Phi_{n_{1},\,n_{2}}%
^{(j)}(\varphi,\,\rho),\;\rho=\frac{eBr^{2}}{2c\hbar},\ \ j=0,1\ ,\nonumber\\
&  \Phi_{n_{1},\,n_{2}}^{(0)}(\varphi,\,\rho)=\mathcal{N}\exp[i(l-l_{0}%
)\varphi]I_{n_{2},n_{1}}\left(  \rho\right)  \;,\ \ n_{1}=m,\ \ n_{2}%
=m-l-\mu,\;-\infty<l\leqslant-1,\ \nonumber\\
&  \Phi_{n_{1},\,n_{2}}^{(1)}(\varphi,\,\rho)=\mathcal{N}\exp[i(l-l_{0}%
)\varphi-i\pi l]I_{n_{1},n_{2}}\left(  \rho\right)  \;,\ \ n_{1}%
=m+l+\mu,\ \ n_{2}=m,\ \ 0\leqslant l\leqslant+\infty\ .\label{2.16}%
\end{align}
Here $l,m\;\left(  m\geq0\right)  $ are two integers, $I_{n,m}(\rho)$ are
Laguerre functions that are related to the Laguerre polynomials $L_{m}%
^{\alpha}(\rho)$ (see eqs. 8.970, 8.972.1 from \cite{GR94}) as follows%
\begin{equation}
I_{m+\alpha,m}(\rho)=\sqrt{\frac{\Gamma\left(  m+1\right)  }{\Gamma\left(
m+\alpha+1\right)  }}e^{-\rho/2}\rho^{\alpha/2}L_{m}^{\alpha}\left(
\rho\right)  ,\ L_{m}^{\alpha}(\rho)=\frac{1}{m!}e^{\rho}\rho^{-\alpha}%
\frac{d^{m}}{d\rho^{m}}e^{-\rho}\rho^{m+\alpha}\,,\label{2.17a}%
\end{equation}
and $\mathcal{N}$ is normalization constant. For any real $\alpha>-1$ the
functions $I_{\alpha+m,\,m}(\rho)$ form a complete orthonormal set on the
half-line $\rho\geqslant0$,%
\begin{equation}
\int_{0}^{\infty}I_{\alpha+k,\,k}(\rho)I_{\alpha+m,\,m}(\rho)d\rho
=\delta_{k,m}\,,\ \ \sum_{m=0}^{\infty}I_{\alpha+m,\,m}(\rho)I_{\alpha
+m,\,m}(\rho^{\prime})=\delta(\rho-\rho^{\prime})\ .\label{f8}%
\end{equation}

Let us define an inner product of two functions $f(\varphi,\,\rho)$
and$\,g(\varphi,\,\rho)$ as
\[
(f,\,g)_{\bot}=\frac{1}{2\pi}\int_{0}^{\infty}d\rho\int_{0}^{2\pi}%
d\varphi\,f^{\ast}(\varphi,\rho)\,g(\varphi,\rho).\
\]
Then eigenfunctions (\ref{2.16}) form an orthogonal set on the $xy$-plane,
\begin{equation}
\left(  \Psi_{n_{1}^{\prime},\,n_{2}^{\prime}}^{(j\,^{\prime})},\Psi
_{n_{1},\,n_{2}}^{(j)}\right)  _{\bot}=\left\vert \mathcal{N}\right\vert
^{2}\,\delta_{n_{1}^{\prime},\,n_{1}}\,\delta_{n_{2}^{\prime},\,n_{2}}%
\,\delta_{j\,^{\prime},\,j}.\label{f23}%
\end{equation}
These functions form a complete orthogonalized set on $D_{H_{\bot}}$.

It is useful to define self-adjoint operators $\hat{R}^{2}$ and$\;R_{c}^{2}$
by analogy with the corresponding classical relations (\ref{17a}) and
(\ref{17}):%
\begin{equation}
\hat{R}^{2}=\frac{2\hat{H}_{\bot}}{M\omega^{2}},\ \hat{R}_{c}^{2}=\hat{R}%
^{2}-\frac{2}{M\omega}\left[  \hat{L}_{z}+\left(  l_{0}+\mu\right)
\hbar\right]  .\label{2.11}%
\end{equation}
In the semiclassical limit the sign of the mean value of the operator $\hat
{R}^{2}-\hat{R}_{c}^{2}$,
\[
\left(  \Psi_{n_{1},\,n_{2}}^{(j\,)},\left(  \hat{R}^{2}-\hat{R}_{c}%
^{2}\right)  \Psi_{n_{1},\,n_{2}}^{(j)}\right)  _{\bot}\left\vert
\mathcal{N}\right\vert ^{-2}=\frac{2\hbar\left(  l+\mu\right)  }{M\omega},
\]
where (\ref{2.15}) is used, allows one to interpret the corresponding states
as particle trajectories that embrace and do not embrace the solenoid. Namely,
an orbit embraces the solenoid for $l\geq0$ (type $j=1$), and do not for
$l\leqslant-1$ (type $j=0$). These classification corresponds to classical one
described in the previous section, see eq. (\ref{17}) and Fig. 1. Trajectories
with $l=0$ and$\;l=-1$ pass most close to the solenoid.

\begin{figure}[ptb]%
\centering
\includegraphics[
height=2.418in,
width=2.9404in
]%
{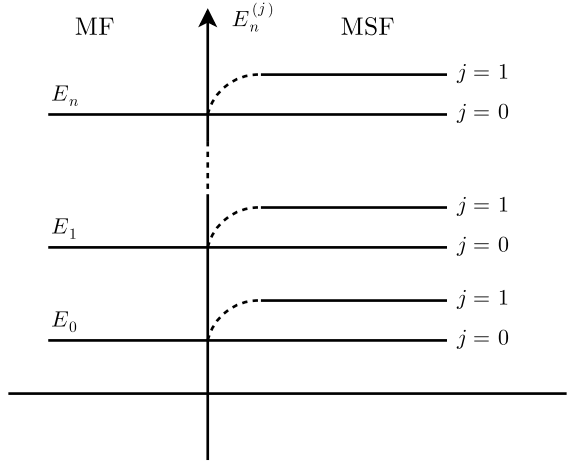}%
\caption{Splitting of Landau levels in MSF.}%
\label{level}%
\end{figure}

If $\mu\neq0$, the degeneracy of energy spectrum is partially lifted, namely,
energy levels of states (\ref{2.16}) with $\;l\geq0$ are shifted with respect
to the Landau levels by $\mu\hbar\omega$, such that $\mathcal{E}_{n_{1}}%
=\hbar\omega\left(  m+l+\mu+1/2\right)  $, while energy levels of states
(\ref{2.16}) with $l\leqslant-1$ are still given by the Landau formula
$\mathcal{E}_{n_{1}}=\hbar\omega\left(  m+1/2\right)  $. For ${\mu}=0$ there
is no any impact of ABS on the energy spectrum. Splitting of the Landau levels
in the MSF is represented on Fig. \ref{level}.%

\section{Coherent states}

\subsection{Instantaneous coherent states on $xy-$plane}

Let us introduce operators $\hat{a}_{1}$, $\hat{a}_{2}$ and $\hat{a}%
_{1}^{\dagger},\hat{a}_{2}^{\dagger}$ that correspond to classical quantities
$a_{1}$, $a_{2}$ and $a_{1}^{\ast}$, $a_{2}^{\ast},$
\begin{align}
\hat{a}_{1}  & =\frac{-i\hat{P}_{x}-\hat{P}_{y}}{\sqrt{2\hbar\,M\omega}%
},\ \ \hat{a}_{2}=\frac{M\omega\,(x+iy)+i\hat{P}_{x}-\hat{P}_{y}}{\sqrt
{2\hbar\,M\omega}};\nonumber\\
\hat{a}_{1}^{\dagger}  & =\frac{i\hat{P}_{x}-\hat{P}_{y}}{\sqrt{2\hbar
\,M\omega}},\ \ \hat{a}_{2}^{\dagger}=\frac{M\omega\,(x-iy)-i\hat{P}_{x}%
-\hat{P}_{y}}{\sqrt{2\hbar\,M\omega}}\ .\label{c1a}%
\end{align}
One ought to say that the momentum operators $\hat{P}_{x}$ and $\hat{P}_{y}$
are symmetric but not self-adjoint on the domain $D_{H_{\bot}}.$ That is why,
one cannot consider $\hat{a}_{1}^{\dagger}$ and $\hat{a}_{2}^{\dagger} $ as
adjoint to $\hat{a}_{1}$ and $\hat{a}_{2}.$ Nevertheless, they play an
important auxiliary role in the further constructions.

Using properties of Laguerre functions, one can find the action of the
operators $\hat{a}_{1}^{\dagger},\,\,\hat{a}_{1};\,\,\hat{a}_{2}^{\dagger
},\,\,\hat{a}_{2}$ on the functions (\ref{2.16}),%
\begin{align}
\hat{a}_{1}\Phi_{n_{1},\,n_{2}}^{(j)}(\varphi,\,\rho)  & =\sqrt{n_{1}}%
\,\Phi_{n_{1}-1,\,n_{2}}^{(j)}(\varphi,\,\rho)\,,\ \ \hat{a}_{1}^{\dagger}%
\Phi_{n_{1},\,n_{2}}^{(j)}(\varphi,\,\rho)=\sqrt{n_{1}+1}\,\Phi_{n_{1}%
+1,\,n_{2}}^{(j)}(\varphi,\,\rho)\,,\nonumber\\
\hat{a}_{2}\Phi_{n_{1},\,n_{2}}^{(j)}(\varphi,\,\rho)  & =\sqrt{n_{2}}%
\,\Phi_{n_{1},\,n_{2}-1}^{(j)}(\varphi,\,\rho)\,,\ \ \hat{a}_{2}^{\dagger}%
\Phi_{n_{1},\,n_{2}}^{(j)}(\varphi,\,\rho)=\sqrt{n_{2}+1}\,\Phi_{n_{1}%
,\,n_{2}+1}^{(j)}(\varphi,\,\rho)\,,\label{f15}%
\end{align}
where possible values of $n_{1}$ and$\ n_{2}$ depend on $m,l,$ and $j$
according to (\ref{2.16}) and the functions $\Phi_{n_{1}+s_{1},\,n_{2}+s_{2}%
}^{(j)}$ are defined as follows%
\begin{align*}
\Phi_{n_{1}+s_{1},\,n_{2}+s_{2}}^{(0)}(\varphi,\,\rho)  & =\mathcal{N}%
\exp[i\epsilon(l_{0}-l-s_{1}+s_{2})\varphi]I_{n_{2}+s_{2},n_{1}+s_{1}}\left(
\rho\right)  ,\ \\
\Phi_{n_{1}+s_{1},\,n_{2}+s_{2}}^{(1)}(\varphi,\,\rho)  & =\mathcal{N}%
\exp\left\{  i\epsilon\left[  (l_{0}-l-s_{1}+s_{2})\varphi+\pi\left(
l+s_{1}-s_{2}\right)  \right]  \right\}  I_{n_{1}+s_{1},n_{2}+s_{2}}\left(
\rho\right)  .
\end{align*}
at $s_{1}=0,\pm1$ and $s_{2}=0,\pm1$. There appear new functions $\Phi
_{n_{1},\,n_{2}-1}^{(0)}(\varphi,\,\rho)$ with $n_{2}=m+1-\mu$ and
$\Phi_{n_{1}-1,\,n_{2}}^{(1)}(\varphi,\,\rho)$ with $n_{1}=m+\mu$, which are
irregular at $r\rightarrow0$. Such functions were not defined by eqs.
(\ref{2.16}). In addition, for\emph{\ }$n_{1}=0$\emph{\ }or\emph{\ }$n_{2}=0$,
one has to bear in mind that%
\[
\hat{a}_{1}\Phi_{0,\,-l-\mu}^{(0)}(\varphi,\,\rho)=0,\;\hat{a}_{2}\Phi
_{l+\mu,0}^{(1)}(\varphi,\,\rho)=0.
\]

Formal commutators for the operators $\hat{a}_{1}^{\dagger},\hat{a}_{1}$
and$\,\,\hat{a}_{2}^{\dagger},\hat{a}_{2}$ have the form:%
\begin{equation}
\left[  \hat{a}_{1},\hat{a}_{1}^{\dagger}\right]  =1+f,\ \left[  \hat{a}%
_{2},\hat{a}_{2}^{\dagger}\right]  =1-f,\ \left[  \hat{a}_{1},\hat{a}%
_{2}\right]  =f,\ \left[  \hat{a}_{1},\hat{a}_{2}^{\dagger}\right]
=0,\label{c4}%
\end{equation}
with a singular function $f=\Phi\left(  \pi Br\right)  ^{-1}\delta
(r)=2(l_{0}+\mu)\delta(\rho).$ However, one can verify by the help of
(\ref{f15}) that this function gives zero contribution on the domain
$D_{H_{\bot}},$ such that on this domain $\hat{a}_{1}^{\dagger},\hat{a}%
_{2}^{\dagger}$ and $\hat{a}_{1},\hat{a}_{2}$ behave as creation and
annihilation operators. Operators $\hat{R}^{2},R_{c}^{2},\hat{r}^{2},\hat
{H}_{\bot},$ and $\hat{L}_{z}$ can be expressed in terms of the operators
$\hat{a}_{1}^{\dagger},\hat{a}_{1}$ and$\,\,\hat{a}_{2}^{\dagger},\hat{a}_{2}$
as follows:%
\begin{align}
& \hat{R}^{2}=\frac{\hbar}{M\omega}\left(  2\hat{N}_{1}+1\right)  ,\;\hat
{R}_{c}^{2}=\frac{\hbar}{M\omega}\left(  2\hat{N}_{2}+1\right)  ,\ x+iy=\sqrt
{\frac{2\hbar}{M\omega}}\left(  \hat{a}_{2}-\hat{a}_{1}^{\dagger}\right)
,\nonumber\\
& \hat{H}_{\bot}=\hbar\omega\left(  \hat{N}_{1}+1/2\right)  ,\ \frac{1}{\hbar
}\hat{L}_{z}+l_{0}+\mu=\left(  \hat{N}_{1}-\hat{N}_{2}\right)  ,\;\hat{N}%
_{s}=\hat{a}_{s}^{\dagger}\hat{a}_{s},\;s=1,2.\label{RR0}%
\end{align}

The functions $\Phi_{n_{1},\,n_{2}}^{(j)}(\varphi,\,\rho)$ (\ref{2.16}) can be
used to construct the following useful states $\mathbf{\Phi}_{z_{1},\,z_{2}%
}^{(j)}(\varphi,\,\rho)$,%
\begin{equation}
\mathbf{\Phi}_{z_{1},\,z_{2}}^{(j)}(\varphi,\,\rho)=\sum_{l}\mathbf{\Phi
}_{z_{1},\,z_{2}}^{(j)l}(\varphi,\,\rho),\;\mathbf{\Phi}_{z_{1},\,z_{2}%
}^{(j)l}(\varphi,\,\rho)=\sum_{m}\frac{z_{1}^{n_{1}}z_{2}^{n_{2}}\Phi
_{n_{1},\,n_{2}}^{(j)}(\varphi,\,\rho)}{\sqrt{\Gamma(1+n_{1})\,\Gamma
(1+n_{2})}},\label{g1}%
\end{equation}
where $z_{1}$ and $z_{2}$ are complex parameters, possible values of $n_{1}$
and$\ \ n_{2}$ depend on $m,l,$ and $j$ according to (\ref{2.16}), and we set
$\mathcal{N}=1$. We call these states instantaneous coherent states on
$xy$-plane. These states can be expressed via special functions $Y_{\alpha
}(z_{1},z_{2};\rho),$%
\begin{equation}
Y_{\alpha}(z_{1},z_{2};\rho)=\sum_{m=0}^{\infty}\frac{z_{1}^{m}\,z_{2}%
^{m+\alpha}I_{m+\alpha,\,m}(\rho)}{\sqrt{\Gamma(1+m)\Gamma(1+m+\alpha)}%
},\label{g6}%
\end{equation}
as follows:%
\begin{align}
\mathbf{\Phi}_{z_{1},\,z_{2}}^{(0)l}(\varphi,\,\rho)  & =\exp[i\epsilon
(l_{0}-l)\varphi]Y_{-l-\mu}(z_{1},z_{2};\rho),\nonumber\\
\mathbf{\Phi}_{z_{1},\,z_{2}}^{(1)l}(\varphi,\,\rho)  & =\exp\left\{
i\epsilon\left[  (l_{0}-l)\varphi+\pi l\right]  \right\}  Y_{l+\mu}%
(z_{2},z_{1};\rho).\label{g2}%
\end{align}
By the help of the well-known sum,
\[
\sum_{m=0}^{\infty}\frac{z^{m}\,I_{\alpha+m,\,m}(x)}{\sqrt{\Gamma
(1+m)\Gamma(1+\alpha+m)}}=z^{-\frac{\alpha}{2}}\exp\left(  z-\frac{x}%
{2}\right)  \,J_{\alpha}(2\sqrt{xz}\,),
\]
where $J_{\alpha}(x)$ are the Bessel functions of the first kind, one can
obtain the following representation for $Y_{\alpha}(z_{1},z_{2};\rho)$:
\begin{equation}
Y_{\alpha}(z_{1},z_{2};\rho)=\exp\left(  z_{1}z_{2}-\frac{\rho}{2}\right)
\left(  \sqrt{\frac{z_{2}}{z_{1}}}\right)  ^{\alpha}J_{\alpha}(2\sqrt
{z_{1}z_{2}\rho}).\label{g3}%
\end{equation}
Then it follows from (\ref{f15}):%
\begin{equation}
\hat{N}_{k}\mathbf{\Phi}_{z_{1},\,z_{2}}^{(j)}(\varphi,\,\rho)=z_{k}%
\partial_{z_{k}}\mathbf{\Phi}_{z_{1},\,z_{2}}^{(j)}(\varphi,\,\rho
)\;,\;k=1,2;\label{Nz}%
\end{equation}
and
\begin{align}
& a_{1}\mathbf{\Phi}_{z_{1},\,z_{2}}^{(j)}(\varphi,\rho)=z_{1}\left[
\mathbf{\Phi}_{z_{1},\,z_{2}}^{(j)}(\varphi,\,\rho)-\left(  -1\right)
^{j}\mathbf{\Phi}_{z_{1},\,z_{2}}^{(j)-1}(\varphi,\,\rho)\right]  ,\nonumber\\
& a_{2}\mathbf{\Phi}_{z_{1},\,z_{2}}^{(j)}(\varphi,\rho)=z_{2}\left[
\mathbf{\Phi}_{z_{1},\,z_{2}}^{(j)}(\varphi,\,\rho)+\left(  -1\right)
^{j}\mathbf{\Phi}_{z_{1},\,z_{2}}^{(j)0}(\varphi,\,\rho)\right]  .\label{h4a}%
\end{align}

Then, using eqs. 6.615 from \cite{GR94}, we obtain:
\begin{align}
& \left(  \mathbf{\Phi}_{z_{1},\,z_{2}}^{(j)},\mathbf{\Phi}_{z_{1}^{\prime
},\,z_{2}^{\prime}}^{(j^{\prime})}\right)  =\delta_{jj^{\prime}}%
\mathcal{R}^{(j)};\nonumber\\
& \mathcal{R}^{(0)}=Q_{1-\mu}\left(  \sqrt{z_{1}^{\ast}z_{1}^{\prime}}%
,\sqrt{z_{2}^{\ast}z_{2}^{\prime}}\right)  ,\ \mathcal{R}^{(1)}=Q_{\mu}\left(
\sqrt{z_{2}^{\ast}z_{2}^{\prime}},\sqrt{z_{1}^{\ast}z_{1}^{\prime}}\right)
,\ \nonumber\\
& Q_{\alpha}(u,\,v)=Q_{\alpha}^{-}(u,\,v)+\left(  \frac{v}{u}\right)
^{\alpha}I_{\alpha}(2uv),\;Q_{\alpha}^{-}(u,\,v)=\sum_{l=1}^{\infty}\left(
\frac{v}{u}\right)  ^{\alpha+l}I_{\alpha+l}(2uv),\label{z2}%
\end{align}
where $I_{\alpha}(u)$ are the modified Bessel functions of the first kind. We
define the mean value of an operator $\hat{F}$\ in the form
\[
{\overline{(F)}}_{(j)}=\left(  \mathbf{\Phi}_{z_{1},\,z_{2}}^{(j)},\hat
{F}\,\mathbf{\Phi}_{z_{1},\,z_{2}}^{(j)}\right)  \left(  \mathbf{\Phi}%
_{z_{1},\,z_{2}}^{(j)},\mathbf{\Phi}_{z_{1},\,z_{2}}^{(j)}\right)  ^{-1}\,.
\]

Using (\ref{Nz}), one can calculate\emph{\ }the mean values of $\hat{N}_{s}$:%
\begin{equation}
{\overline{(N_{s})}}_{(j)}=z_{s}\left.  \partial_{z_{s}^{\prime}}%
\ln\mathcal{R}^{(j)}\right\vert _{z_{s}^{\prime}=z_{s}},\;s=1,2.\label{meanN}%
\end{equation}
This allows one to connect means values of $\hat{R}^{2}$ and $\hat{R}_{c}^{2}
$ with parameters $z_{1}$ and $z_{2}$. We expect in the semiclassical limit
that ${\overline{(N_{s})}}_{(j)}\approx\left\vert z_{s}\right\vert ^{2}$. At
the same time length scales defined by the means ${\overline{(R^{2})}}_{(j)}$,
${\overline{(R_{c}^{2})}}_{(j)}$ have to be large enough which implies
$\left\vert z_{s}\right\vert ^{2}\gg1$. We expect that the sign of the
difference ${\overline{(R^{2})}}_{(j)}-{\overline{(R_{c}^{2})}}_{(j)}$ is
related to the trajectory type if the difference is sufficiently large, such
that for states with $j=0$ we have $\left\vert z_{1}\right\vert ^{2}%
<\left\vert z_{2}\right\vert ^{2}$, and for states with $j=1$, we have
$\left\vert z_{1}\right\vert ^{2}>\left\vert z_{2}\right\vert ^{2}$. We note
that in both cases the corresponding functions $Q_{\alpha}(u,v)$ are
calculated at $\left\vert v\right\vert >\left\vert u\right\vert \gg1$.

There exist all the derivatives $\partial_{v}\left[  \left(  v/u\right)
^{\alpha+l}I_{\alpha+l}(2uv)\right]  $, the series $Q_{\alpha}^{-}(u,\,v)$
converges and the series of derivatives $\sum_{l=1}^{\infty}\partial
_{v}\left[  \left(  v/u\right)  ^{\alpha+l}I_{\alpha+l}(2uv)\right]  $
converges uniformly on the half-line, $0<\operatorname{Re}v<\infty$. Thus, one
arrives to a differential equation with respect to $Q_{\alpha}^{-}(u,\,v)$,
\[
\frac{dQ_{\alpha}^{-}(u,\,v)}{dv}=2v\left[  \left(  v/u\right)  ^{\alpha
}I_{\alpha}(2uv)+Q_{\alpha}^{-}(u,\,v)\right]  .
\]
To evaluate asymptotics, we represent its solution as follows:
\begin{equation}
Q_{\alpha}^{-}(u,\,v)=e^{u^{2}+v^{2}}\left[  1-T(u,\,v)\right]
,\;T(u,\,v)=2e^{-u^{2}}\int_{v}^{\infty}e^{-\tilde{v}^{2}}\left(  \frac
{\tilde{v}}{u}\right)  ^{\alpha}I_{\alpha}(2u\tilde{v})\tilde{v}d\tilde
{v},\label{z4}%
\end{equation}
where formula 6.631.4 \cite{GR94} is used. Then
\begin{equation}
Q_{\alpha}(u,\,v)=e^{u^{2}+v^{2}}\tilde{Q}_{\alpha}(u,\,v),\;\tilde{Q}%
_{\alpha}(u,\,v)=\left[  1-T(u,\,v)+e^{-u^{2}-v^{2}}\left(  v/u\right)
^{\alpha}I_{\alpha}(2uv)\right]  .\label{z5}%
\end{equation}
Thus, the mean values (\ref{meanN}) have the form:%
\begin{align}
& {\overline{(N_{s})}}_{(j)}=\left\vert z_{s}\right\vert ^{2}+z_{s}\left.
\partial_{z_{s}^{\prime}}\ln\mathcal{\tilde{R}}^{(j)}\right\vert
_{z_{s}^{\prime}=z_{s}},\;s=1,2,\nonumber\\
& \mathcal{\tilde{R}}^{(0)}=\tilde{Q}_{1-\mu}\left(  \sqrt{z_{1}^{\ast}%
z_{1}^{\prime}},\sqrt{z_{2}^{\ast}z_{2}^{\prime}}\right)  ,\;\mathcal{\tilde
{R}}^{(1)}=\tilde{Q}_{\mu}\left(  \sqrt{z_{2}^{\ast}z_{2}^{\prime}}%
,\sqrt{z_{1}^{\ast}z_{1}^{\prime}}\right)  .\ \label{z6}%
\end{align}
Using asymptotics of the function $I_{\alpha}(2uv)$, one can verify that if
$\left\vert v\right\vert >\left\vert u\right\vert \gg1$ then $\left\vert
z_{s}\right\vert ^{2}\gg z_{s}\left.  \partial_{z_{s}^{\prime}}\ln
\mathcal{\tilde{R}}^{(j)}\right\vert _{z_{s}^{\prime}=z_{s}}$ in (\ref{z6}).
For semiclassical states corresponding to orbits placed far enough from the
solenoid, i.e., for $\left\vert \left\vert z_{1}\right\vert ^{2}-\left\vert
z_{2}\right\vert ^{2}\right\vert $ $\gg1$, the contribution $z_{s}\left.
\partial_{z_{s}^{\prime}}\ln\mathcal{\tilde{R}}^{(j)}\right\vert
_{z_{s}^{\prime}=z_{s}}$ is small as $\exp\left(  -\left\vert \left\vert
z_{1}\right\vert ^{2}-\left\vert z_{2}\right\vert ^{2}\right\vert \right)  $.
Finally, we obtain:%
\begin{equation}
\left\vert z_{1}\right\vert ^{2}\approx\frac{M\omega}{2\hbar}{\overline
{(R^{2})}}_{(j)},\;\;\left\vert z_{2}\right\vert ^{2}\approx\frac{M\omega
}{2\hbar}{\overline{(R_{c}^{2})}}_{(j)},\ \ \left\vert z_{s}\right\vert
^{2}\gg1.\label{z7}%
\end{equation}

By the help of (\ref{h4a}), one can find:%
\begin{align}
{\overline{(a_{1})}}_{(0)}  & =z_{1}\Delta_{1-\mu}(|z_{1}|,|z_{2}%
|),\ \ {\overline{(a_{2})}}_{(0)}=z_{2},\ \ {\overline{(a_{1})}}_{(1)}%
=z_{1},\nonumber\\
{\overline{(a_{2})}}_{(1)}  & =z_{2}\Delta_{\mu}(|z_{2}|,|z_{1}|),\ \ \Delta
_{\alpha}(u,v)=\frac{Q_{\alpha}^{-}(u,\,v)}{Q_{\alpha}(u,\,v)},\label{z8}%
\end{align}
such that these means match with eqs. (\ref{z6}) in the classical limit.

\subsection{Time-dependent coherent states}

Consider Schrödinger equation with the complete three-dimensional Hamiltonian
$\hat{H}$ (\ref{H}) and corresponding solutions $\Psi
(t,\,\mbox{\boldmath$r$})$ with a given momentum $p_{z}$,%
\[
\Psi(t,\,\mbox{\boldmath$r$})=\mathcal{N}\exp\left\{  -\frac{i}{\hbar}\left[
\left(  \frac{p_{z}^{2}}{2M}+\frac{\hbar\omega}{2}\right)  t-p_{z}z\right]
\right\}  \Phi(t,\,\varphi,\,\rho),
\]
where $\mathcal{N}$ is normalization constant. The functions $\Phi
(t,\,\varphi,\,\rho)$ obey the following equation
\begin{equation}
i\partial_{t}\Phi(t,\,\varphi,\,\rho)=\omega\hat{N}_{1}\Phi(t,\,\varphi
,\,\rho)\ .\label{l17}%
\end{equation}
One can obey (\ref{l17}) setting $\Phi(t,\,\varphi,\,\rho)=\left.
\mathbf{\Phi}_{z_{1},\,z_{2}}^{(j)}(\varphi,\,\rho)\right\vert _{z_{1}%
=z_{1}\left(  t\right)  },$ where $z_{1}\left(  t\right)  $ is a complex
function of time $t$. Then
\begin{equation}
i\partial_{t}\Psi_{z_{1},\,z_{2}}^{(j)}=i{\dot{z}}_{1}\partial_{z_{1}%
}\mathbf{\Phi}_{z_{1},\,z_{2}}^{(j)}\ ,\ \ \dot{z}_{1}=dz_{1}%
/dt\ .\label{l18a}%
\end{equation}
Substituting (\ref{l18a}) into (\ref{l17}), we find $i{\dot{z}}_{1}=\omega
z_{1}$, where (\ref{Nz}) is used. It is convenient to write a solution for
$z_{1}(t)$ as follows:
\begin{equation}
z_{1}(t)=-|z_{1}|\exp(-i\psi),\ \ \psi=\omega t+\psi_{0},\label{119}%
\end{equation}
where $|z_{1}|$ is\ a given constant. Thus the functions
\begin{equation}
\Psi_{\mathrm{CS}}^{(j)}(t,\,\mbox{\boldmath$r$})=\mathcal{N}\exp\left\{
-\frac{i}{\hbar}\left[  \left(  \frac{p_{z}^{2}}{2M}+\frac{\hbar\omega}%
{2}\right)  t-p_{z}z\right]  \right\}  \mathbf{\Phi}_{z_{1}(t),\,z_{2}}%
^{(j)}(\varphi,\,\rho)\label{l21}%
\end{equation}
are solutions of the Schrödinger equation. At the same time they have special
properties that allow us to treat them as coherent (and under certain
conditions as semiclassical) states.

Let us consider mean values ${\overline{(x)}}_{(j)}$ and ${\overline{(y)}%
}_{(j)}$ of the coordinates with respect to the states $\Psi_{\mathrm{CS}%
}^{(j)}$. To this end it is enough to find the mean value ${\overline{(x+iy)}%
}_{(j)}$. By the help of (\ref{RR0}) we obtain:%
\[
{\overline{(x+iy)}}_{(j)}=\sqrt{\frac{2\hbar}{M\omega}}\left[  {\overline
{(a_{2})}}_{(j)}-{\overline{(a_{1})}}_{(j)}^{\ast}\right]  .
\]
Taking into account eqs. (\ref{z8}) and (\ref{119}), one can see that a point
with coordinates ${\overline{(x)}}_{(j)}$ and ${\overline{(y)}}_{(j)}$ is
moving along a circle on the $xy$-plane with the cyclotron frequency $\omega$,
i.e., its trajectory has the classical form. The same equations allows one to
find a radius ${\overline{(R)}}_{(j)}$ of such a circle and the distance
${\overline{(R_{c})}}_{(j)}$ between its center and the origin,%
\begin{align*}
{\overline{(R)}}_{(0)}  & =\sqrt{\frac{2\hbar}{M\omega}}\left\vert
z_{1}\right\vert \Delta_{1-\mu}(|z_{1}|,|z_{2}|),\ \ {\overline{(R_{c})}%
}_{(0)}=\sqrt{\frac{2\hbar}{M\omega}}\left\vert z_{2}\right\vert ;\\
{\overline{(R)}}_{(1)}  & =\sqrt{\frac{2\hbar}{M\omega}}\left\vert
z_{1}\right\vert ,\ \ {\overline{(R_{c})}}_{(1)}=\sqrt{\frac{2\hbar}{M\omega}%
}\left\vert z_{2}\right\vert \Delta_{\mu}(|z_{2}|,|z_{1}|).
\end{align*}
However, in the general case, the quantities ${\overline{(R)}}_{(j)}$ and
${\overline{(R_{c})}}_{(j)}$ do not coincide with the corresponding
quantities
\[
\sqrt{{\overline{(R^{2})}}_{(j)}}=\sqrt{\frac{\hbar}{M\omega}}\sqrt
{2{\overline{(N_{1})}}_{(j)}+1},\;\sqrt{{\overline{(R_{c}^{2})}}_{(j)}}%
=\sqrt{\frac{\hbar}{M\omega}}\sqrt{2{\overline{(N_{2})}}_{(j)}+1},\;
\]
which are expressed in terms of mean values of the operators $\hat{H}_{\bot} $
and $\hat{L}_{z}$ according to (\ref{RR0}), see also (\ref{meanN}).

It follows from eq. (\ref{z8}) that $\Delta_{1-\mu}(|z_{1}|,|z_{2}|)<1$\ and
$\Delta_{\mu}(|z_{2}|,|z_{1}|)<1$. This allows us to give the following
interpretation for two types of states with $j=0,1.$ States with $j=1$
correspond to orbits that embrace the ABS (which corresponds to $\left\vert
z_{1}\right\vert ^{2}\gtrsim\left\vert z_{2}\right\vert ^{2}$ in the
semiclassical limit). For such orbits $\overline{(R_{c})}_{(1)}<R_{c},$ where
the quantity $R_{c}=\sqrt{2\hbar/M\omega}\left\vert z_{2}\right\vert $ is
interpreted by us as a distance between ABS and the orbit center as a
consequence of eq. (\ref{z7}). At the same time, the mean radius of the orbit
coincides with the classical radius $R=\sqrt{2\hbar/M\omega}\left\vert
z_{1}\right\vert $. The interpretation of $R$ as the classical radius follows
from eq. (\ref{z7}). States with $j=0$ correspond to orbits that do not
embraces the ABS\ (which corresponds to $\left\vert z_{1}\right\vert
^{2}\lesssim\left\vert z_{2}\right\vert ^{2}$ in the semiclassical limit). For
such orbits $\overline{(R_{c})}_{(0)}=R_{c}$ and $\overline{(R)}_{(0)}<R$.

\begin{figure}[ptb]%
\centering
\includegraphics[
height=2.7726in,
width=2.8712in
]%
{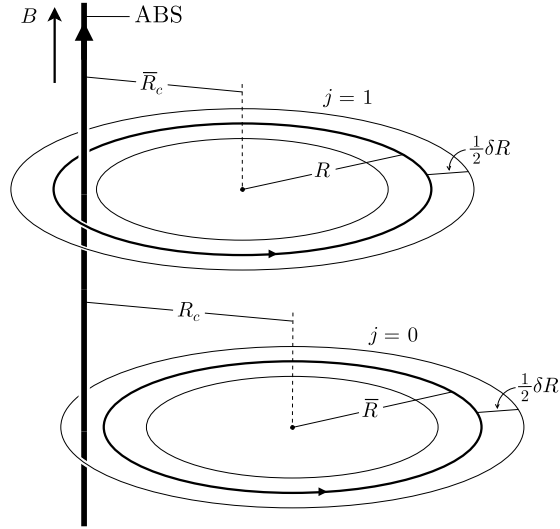}%
\caption{ Spread of particle position around two types of semiclassical orbits
at $R\approx R_{c} $, where $\overline{R}_{c}=\overline{(R_{c})}_{(1)}$ and
$\overline{R}=\overline{(R)}_{(0)}$.}%
\label{spread}%
\end{figure}

By using formulas (\ref{z5}), (\ref{z6}), (\ref{z8}) one can calculate the
variances for $\hat{R}^{2}$, $\hat{R}_{c}^{2}$, and $x+y$ with respect to the
coherent states in the semiclassical limit. With this result one can see these
variances are relatively small for the semiclassical orbits situated far
enough from the solenoid, i.e., for $\left\vert \left\vert z_{1}\right\vert
^{2}-\left\vert z_{2}\right\vert ^{2}\right\vert \gg1$, In this case the
coherent states are highly concentrated around the classical orbits. In the
most interesting case when a semiclassical orbit is situated near the
solenoid, such that the condition $\left\vert \left\vert z_{1}\right\vert
^{2}-\left\vert z_{2}\right\vert ^{2}\right\vert \ll1$ holds, the variance for
$x+y$ increases significantly while the variances for $\hat{R}^{2}$, $\hat
{R}_{c}^{2}$ remain relatively small. In this case $R\approx R_{c}$, however,
one has $\overline{(R_{c})}_{(1)}<R$ and $\overline{(R)}_{(0)}<R_{c}$, as of
course it must be for such semiclassical orbits. Having in mind that the
standard deviation of $x+y$, $\delta R$, is relatively large at $R\approx
R_{c}$, such that $\delta R\gg\left\vert R-\overline{(R_{c})}_{(1)}\right\vert
,\left\vert \overline{(R)}_{(0)}-R_{c}\right\vert $, we illustrate the typical
spread of particle position around two types of semiclassical orbits at
$R\approx R_{c}$ on Fig. \ref{spread}.

Thus, for $\mu\neq0,$ classical relations between parameters of particle
trajectory in constant magnetic field, such that relations between circle
parameters\textrm{\ }$R$ (related to particle energy) and $R_{c}$ (related to
particle angular momentum) are affected in the presence of ABS. Such relations
do not feel the presence of ABS for $\mu=0$, and, even for $\mu\neq0,$ in the
classical limit (in the leading approximation for sufficiently large radii)
discussed above.

Thus, in contrast to the problem in the constant uniform magnetic field (and
in contrast to any problem with quadratic Hamiltonian) in the
magnetic-solenoid field, we meet a completely new situation. Here
time-dependent coherent states can be constructed (which is completely
nontrivial fact due to nonquadratic nature of the Hamiltonian in the
magnetic-solenoid field), the respective mean values move along\emph{\ }%
classical trajectories, however classical relations between physical
quantities imply additional semiclassical restrictions. Not all coherent
states correspond to a semiclassical approximation, which is natural for
nonquadratic Hamiltonians.

Finally, we ought to mention that only linear combinations of the form
\[
\Psi(c_{0},c_{1};t,\,\mbox{\boldmath$r$})=c_{0}\Psi_{\mathrm{CS}}%
^{(0)}(t,\,\mbox{\boldmath$r$})+c_{1}\Psi_{\mathrm{CS}}^{(1)}%
(t,\,\mbox{\boldmath$r$}),
\]
with $c_{0}$ and $c_{1}$ -arbitrary and $c_{0}c_{1}\neq0$ were considered
earlier as coherent states in \cite{CS-old}. Mean values of the operators
$\hat{a}_{1}$ and $\hat{a}_{2}$ in such mixed states do not coincide with the
classical expressions (\ref{l9}).

\subparagraph{{\protect\large Acknowledgement}}

V. G. Bagrov: this work is partially supported by Russian Science and
Innovations Federal Agency under contract No 02.740.11.0238 and Russia
President grant SS-3400.2010.2; S.P. Gavrilov thanks FAPESP for support and
Universidade de São Paulo for hospitality; D.M. Gitman acknowledges the
permanent support of FAPESP and CNPq; D. P. Meira Filho thanks CNPq for a support.

\end{document}